\documentclass[acmsmall]{acmart}

\AtBeginDocument{%
  }
\usepackage{graphicx}
\usepackage{subcaption}
\usepackage{xcolor}
\usepackage{colortbl}
\usepackage{diagbox} 
\usepackage{tabularx}
\usepackage{xcolor,soul,framed} %,caption
\usepackage{multirow,diagbox}
\colorlet{shadecolor}{yellow}
\usepackage{algorithm}
\usepackage{algorithmic}
\usepackage{amsmath}
\usepackage{float}
\usepackage{gensymb}
\usepackage{wasysym}

\setcopyright{acmlicensed}
\copyrightyear{2018}
\acmYear{2018}
\acmDOI{XXXXXXX.XXXXXXX}

\acmJournal{JACM}
\acmVolume{37}
\acmNumber{4}
\acmArticle{111}
\acmMonth{8}

\begin{document}

\title{M2CVD: Enhancing Vulnerability Semantic through Multi-Model Collaboration for Code Vulnerability Detection}


% \author{Ziliang Wang,Ge Li*,Jia Li and Jia Li}
\author{Ziliang Wang}
\email{wangziliang@pku.edu.cn}
\author{Ge Li}
\authornote{Corresponding author.}
\email{geli@pku.edu.c}
\author{Jia Li $\male$}
\email{lijia@stu.pku.edu.cn}
\author{Jia Li}
\email{lijiaa@pku.edu.cn}
\affiliation{%
  \institution{Key Lab of High Confidence Software Technology, MoE, School
of Computer Science, Peking University}
  \city{BeiJing}
  \country{China}
}

% \author{Ge Li}
% \authornote{Corresponding author.}
% \affiliation{%
%   \institution{Peking University}
% %   \streetaddress{1 Th{\o}rv{\"a}ld Circle}
%   \city{BeiJing}
% \email{lige@pku.edu.cn}

% \author{Jia Li \♂}
% %\authornotemark[1]
% % \authornote{Corresponding author.}
% %\orcid{0000-0002-8645-0680}
% \affiliation{%
%   \institution{Peking University}
% %   \streetaddress{1 Th{\o}rv{\"a}ld Circle}
%   \city{BeiJing}
% \email{lijia@stu.pku.edu.cn}

% \author{Yingfei Xiong}
% \affiliation{%
%   \institution{Peking University}
% %   \streetaddress{1 Th{\o}rv{\"a}ld Circle}
%   \city{BeiJing}
% \email{xiongy@pku.edu.cn}

% \author{Jia Li}
% \affiliation{%
%   \institution{Peking University}
% %   \streetaddress{1 Th{\o}rv{\"a}ld Circle}
%   \city{BeiJing}
% \email{lijiaa@pku.edu.cn}

\author{Meng Yan}
\affiliation{%
  \institution{Chongqing University}
  \city{Chongqing}
  \country{China}}
\email{mengy@cqu.edu.cn}

\author{Yingfei Xiong}
\email{xiongyf@pku.edu.cn}
\author{Zhi Jin}
% \authornote{Corresponding author.}
\email{zhijin@sei.pku.edu.cn}

\affiliation{%
  \institution{Key Lab of High Confidence Software Technology, MoE, School
of Computer Science, Peking University}
  \city{BeiJing}
  \country{China}
}

% \author{Yingfei Xiong and Zhi Jin}
% \affiliation{%
%   \institution{Key Lab of High Confidence Software Technology, MoE, School
% of Computer Science, Peking University}
% %   \streetaddress{1 Th{\o}rv{\"a}ld Circle}
%   \city{BeiJing}
% \email{zhijin@sei.pku.edu.cn}

\renewcommand{\shortauthors}{Ziliang Wang,Ge Li et al.}

\begin{abstract}
Large Language Models (LLMs) have strong capabilities in code comprehension, but fine-tuning costs and semantic alignment issues limit their project-specific optimization; conversely, code models such as CodeBERT are easy to fine-tune, but it is often difficult to learn vulnerability semantics from complex code languages.
To address these challenges, this paper introduces the Multi-Model Collaborative Vulnerability Detection approach (M2CVD) that leverages the strong capability of analyzing vulnerability semantics from LLMs to improve the detection accuracy of code models.
M2CVD employs a novel collaborative process: first enhancing the quality of vulnerability semantic description produced by LLMs through the understanding of project code by code models, and then using these improved vulnerability semantic descriptions to boost the detection accuracy of code models.
M2CVD include three main phases:
1) Initial Vulnerability Detection: 
The initial vulnerability detection is conducted by fine-tuning a detection model (e.g., CodeBERT) and interacting with a LLM (e.g., ChatGPT) respectively. The vulnerability description will be generated by the LLM when the code is detected vulnerable by the LLM.
2) Vulnerability Description Refinement: By informing the LLM of the vulnerability assessment results of the detection model, we refine the vulnerability description by interacting with the LLM. Such refinement can enhance LLM's vulnerability understanding in specific projects, effectively bridging the previously mentioned alignment gap; 
3) Integrated Vulnerability Detection: M2CVD integrates code fragment and the refined vulnerability descriptions inferred to form synthetic data.
Then, the synthetic data is used to fine-tune a validation model, optimize the defect feature learning efficiency of the model, and improve the detection accuracy.
We demonstrated M2CVD's effectiveness on two real-world datasets, where M2CVD significantly outperformed the baseline. 
In addition, we demonstrate that the M2CVD collaborative method can extend to other different LLMs and code models to improve their accuracy in vulnerability detection tasks.
\end{abstract}





\maketitle

\section{Introduction}

Vulnerabilities in software refer to code weaknesses that can be easily exploited, which can lead to serious consequences such as unauthorized information disclosure ~\cite{fu2022linevul} and cyber extortion ~\cite{thapa2022transformer}. 
Recent statistics underscore this burgeoning issue: In Q1 of 2022, the US National Vulnerability Database (NVD) disclosed 8,051 vulnerabilities, marking a 25\% increase from the previous year ~\cite{cheng2022path}.
Further accentuating this trend, a study revealed that out of 2,409 analyzed codebases, 81\% had at least one recognized open-source vulnerability.
The increasing scale and ubiquity of these vulnerabilities emphasize the need for well-developed automated vulnerability detection mechanisms.
Such a detection system helps to strengthen software security and forestall a range of potential security risks~\cite{fu2022linevul,thapa2022transformer,jang2014survey,johnson2011guide}.

The vulnerability detection models in the existing literature are mainly divided into two categories: (1) conventional detection models ~\cite{yamaguchi2015pattern,yamaguchi2017pattern} and (2) Deep Learning (DL)-based models ~\cite{li2021vulnerability,duan2019vulsniper,russell2018automated,li2018vuldeepecker,dam2017automatic}
The former typically requires experts to manually formulate detection rules~\cite{Checkmarx,Flawfinder}.
These methods are usually labor-intensive to create and are difficult to achieve low false positive rates and low false negative rates ~\cite{li2018vuldeepecker,li2021sysevr}. 
On the contrary, deep learning (DL) -based detection methods %\yxmodifyok{utilize data mining and machine learning paradigms for vulnerability prediction}
learn the patterns of vulnerabilities from a training set~\cite{li2018vuldeepecker,dam2017automatic,lin2017poster}.
They avoid manual heuristic methods and autonomously learn and identify vulnerability features.
In order to further learn the semantics of vulnerabilities, methods based on pre-trained code models~\cite{nguyen2022regvd} and vulnerability detection studies based on LLMs~\cite{fu2023chatgpt} have been proposed successively.
In summary, traditional vulnerability detection methods usually depend on pre-defined rules, a process of expert interventions, rendering them laborious and occasionally imprecise. 
The DL-based detection methods, by contrast, show better detection ability by learning vulnerable code patterns automatically.

In the latest research~\cite{steenhoek2023empirical}, the efficacy of pre-trained language models for software vulnerability detection has been extensively explored, encompassing LLMs such as ChatGPT~\cite{ChatGPT} and LLaMa~\cite{Llama}, alongside pre-trained code models like CodeBERT~\cite{feng2020codebert} and UniXcoder~\cite{guo2022unixcoder}. 
Compared with traditional deep learning networks, these models show more excellent performance in code vulnerability detection tasks after fine-tuning~\cite{nguyen2022regvd}. Figure 1 illustrates the flow of code model fine-tuning.
Detecting vulnerabilities using pre-trained models has its benefits, but when it comes to real-world applications, we encounter the first challenge that \textbf{the complexity of code makes it hard for the code model to learn vulnerability semantics}~\cite{steenhoek2023language}. 
%
% \textbf{Code models struggle to grasp the features of vulnerabilities}~\cite{steenhoek2023language}. 
%
% \yxmodify{Small-scale code models like CodeBERT and UniXcoder can be fine-tuned on a dataset from a specific area, but this dataset might only have labels to show if a piece of code is vulnerable or not. 
% %
% This makes it difficult for the model to understand why a piece of code is vulnerable.
% Moreover, since the pre-training datasets for code models usually do not have vulnerability descriptions, so these models tend to learn code patterns(e.g,. if-else,for) and miss the actual reasons (e.g., memory allocation) for vulnerabilities~\cite{steenhoek2023empirical}.}
% {
The pre-training datasets for code models usually do not have vulnerability descriptions. Though code models can be fine-tuned on a domain-specific vulnerability dataset, these datasets usually only contain labels to show if a piece of code is vulnerable or not. Without semantic description, it would be difficult for code models to learn the actual cause of the vulnerabilities. For the same reason, existing vulnerability detection methods usually only output a vulnerability judgment indicating whether the code is vulnerable. 
% }

In contrast to existing approaches that use pattern matching to enhance vulnerability semantic ~\cite{steenhoek2023language}, we resort to the strong understanding abilities of LLMs to create natural language descriptions of vulnerable code, so as to make connections between code and the causes of vulnerabilities. This will bring two benefits: the more abstract natural language description will help the code model better learn the semantics of the vulnerability, and the vulnerability description can help programmers better determine the cause of the vulnerability to maintain the code.
In the latest research, this process is exploited for natural language understanding\cite{min2024synergetic}.
But we've hit the second challenge, \textbf{the semantic alignment problem of LLMs}.
Given the scale of LLMs, fine-tuning them on a specific domain is challenging.
Real projects, organizations, or specific fields have their own coding rules and business logic. 
Using LLMs trained on data from open domain might not make accurate vulnerability judgment on code in a specific domain (e.g., \citet{fu2023chatgpt} reports a F1 score of 29\% with GPT4). 
%
% \yxmodify{
As a result, LLMs may generate incorrect vulnerability description.

\begin{figure*}
\centering
\includegraphics[width=0.98\textwidth]{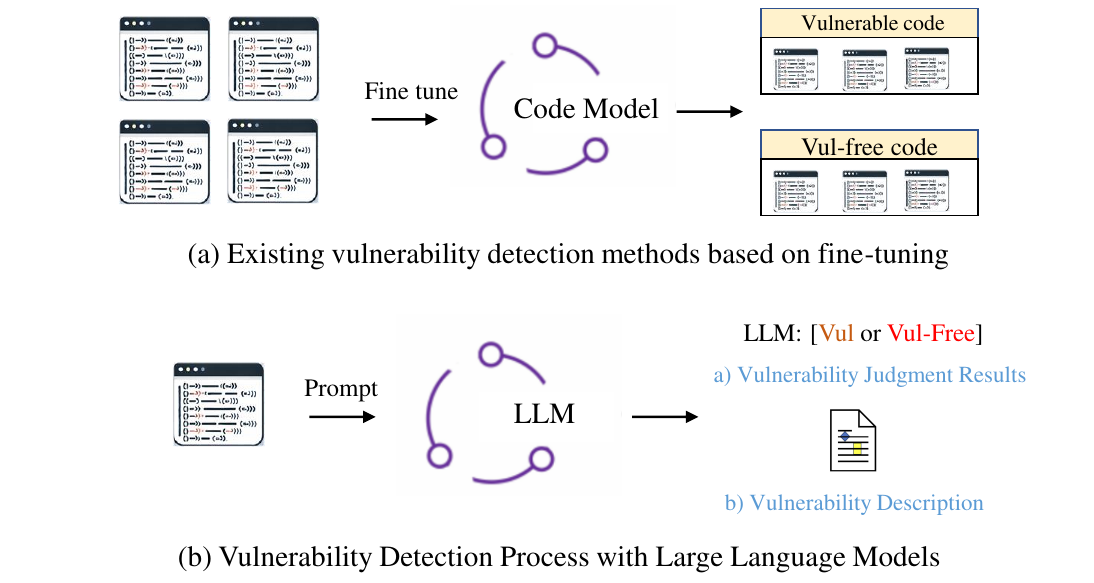}
% \captionsetup{justification=centering}
\caption{Existing vulnerability detection method processes based on code model fine-tuning and interactive vulnerability detection processes based on large language models.
% \yxcomment{The inverse of vulnerable is invulnerable but not un-vulnerable.} 
}
\label{fig:main}
\end{figure*}

In this paper, we have proposed M2CVD, an innovative approach that combines the strengths of pre-trained code models and LLMs to better detect vulnerabilities.
For the first challenge, we rely on the ability of LLMs to interpret vulnerabilities, leveraging the explanatory text to help code models to understand the semantics of vulnerabilities.
%
% For the first challenge, we fine-tune a code model with a dataset of vulnerabilities from a specific field. The results from this fine-tuned model will be used to help improve the LLM's results.
%
For the second challenge, we rely on the advantage of code models that is easy to fine-tune, and use their judgment results to enhance the vulnerability semantic understanding of LLMs for specific projects.
In this way, M2CVD can help operators to improve the accuracy of vulnerability detection through the collaborative interaction process combined with LLMs API without changing the existing code model structure.
In summary, 
the main contributions of this paper are as follows:
\begin{itemize}
% \vspace{-1em}
\item [a)]We propose M2CVD, an approach that integrates the capabilities of pre-trained code models and LLMs to better utilize their strengths for enhancing the precision of vulnerability detection tasks. 
Compared with the existing vulnerability detection, M2CVD supports the output of vulnerability semantic description to assist programmers to maintain code.
% \vspace{-1em}
\item [b)]This paper proposes a vulnerability semantic description refinement method, which leverages the insights of fine-tuning pre-trained code models on specific data to effectively enhance the vulnerability description generation ability of unfine-tuned LLMs on project-specific domain code.
% \vspace{-1em}
\item [c)] We evaluate our approach through extensive experimentation on two real-world datasets. Experimental results show that the M2CVD can still improve the performance of code vulnerability detection with the different of pre-trained code model and LLMs.
\end{itemize}

\textbf{Data Availability.} We open-source our replication package\footnote{Our  replication package (data and code) :https://github.com/HotFrom/M2CVD}, including the datasets and the source code of M2CVD, to facilitate other researchers and practitioners to repeat our work and verify their studies. 

\textbf{Paper Organization.} Section 2 describes the background of code vulnerability detection. Section 3 presents our model M2CVD. Sections 4 and 5 describe the datasets and experiments of our study, respectively. Sections 6  discuss a case of M2CVD, respectively. Section 7 concludes the discussion of M2CVD. And Section 8 includes summary of our approach and future directions.

\section{Related work}

\subsection{Traditional Vulnerability Detection}
Over the years, a lot of methods for vulnerability detection have emerged. Overall, initial research in this area focused on identifying vulnerabilities by means of manually customized rules ~\cite{Checkmarx,Flawfinder}. 
While these approaches provide heuristic approaches to vulnerability  detection, they require extensive manual analysis and formulation of defect patterns.
In addition, syntactic elements are repeated in different code fragments, as prescribed by certain rules, have been observed to induce elevated rates of both false positives and false negatives ~\cite{yamaguchi2015pattern,yamaguchi2017pattern,li2021sysevr}.

\subsection{Deep Neural Network for Vulnerability Detection}
To minimize human intervention, recent works have turned to employing neural network-based models for the extraction of vulnerability features from code fragments ~\cite{dam2017automatic,russell2018automated}.
Existing deep learning-based vulnerability detection models predominantly bifurcate into two classifications: token-based and graph-based models.

Token-based models treat code as a linear sequence and use neural networks to learn vulnerability features from known cases, aiming to identify previously undetected vulnerabilities. ~\cite{russell2018automated,li2018vuldeepecker,cheng2021deepwukong}. 
For instance, Russell et al. harnessed the power of both recurrent neural networks (RNNs) and convolutional neural networks (CNNs) to learn feature sets from code token sequences tailored for vulnerability identification~\cite{russell2018automated}.
Concurrently, Li et al. ~\cite{li2018vuldeepecker} employed BiLSTM ~\cite{schuster1997bidirectional} to encode a segmented version of input code,  known as  'code gadget', centered on key markers, especially  library/API function calls. 
However, these token-based models often ignore the complexity of the source code structure, which may lead to inaccurate detection.

While focusing on token-based models, another research direction is to reveal the potential of graph-based methods in the field of vulnerability detection~\cite{li2021vulnerability,chakraborty2021deep,zheng2021vu1spg,nguyen2022regvd}.
DeepWukong ~\cite{cheng2021deepwukong} utilizes GNN for feature learning, which focuses on compressing code fragments into a dense, low-dimensional vector space to enhance the detection of a large number of vulnerability types.
DeepTective ~\cite{rabheru2021deeptective} confronts vulnerabilities common to PHP scripts such as SQL injection, Cross-Site Scripting (XSS), and command injection by deploying a combination of Gated Recurrent Units (GRUs) and Graph Convolutional Networks.
%
% An amalgamated approach is further examined in \cite{yan2018new}, where the conception of a Hybrid Deep Learning Network is elucidated, aimed specifically at thwarting code injection attacks within the ambit of HTML5-based applications.

Graph-based detection models learn code structure through varied graph representations, utilizing neural networks for vulnerability detection~\cite{wu2022vulcnn,cao2022mvd}.
For instance, Zhou et al. ~\cite{zhou2019devign} used the gated graph recurrent network ~\cite{li2016gated}, extracting structural details from triadic graph representations—AST, CFG, and DFG. 
Chakraborty et al. ~\cite{chakraborty2021deep} introduced REVEAL, an innovative approach that amalgamation of the gated graph neural network, re-sampling techniques ~\cite{chawla2002smote}, and triplet loss ~\cite{mao2019metric}.
Wu et al. ~\cite{wu2022vulcnn} proposed a approach that can efficiently convert the source code of a function into an image while preserving the program details.
Meanwhile, Cao et al. ~\cite{cao2022mvd} proposed a statement-centric approach, based on flow-sensitive graph neural network, to understanding semantic and structural data.

\subsection{Pre-Trained Models for Vulnerability Detection}

Taking inspiration from the success of pre-trained models in the field of natural language processing (NLP), an increase of related research works aims to leverage these pre-trained models to improve code vulnerability detection accuracy ~\cite{feng2020codebert,kanade2020learning,niu2022spt,lin2021traceability,bai2021syntax,guo2022unixcoder}. 

The core idea of these works is a pre-trained model on a large amount of source code data, followed by specialized fine-tuning for a specific task. ~\cite{kanade2020learning}. 
To illustrate, Feng et al. ~\cite{feng2020codebert} proposed CodeBERT specifically for understanding and generating source code, which combines the processing power of natural and programming languages.
Similarly, CuBERT combines masked language modeling with sentence prediction for code representation ~\cite{kanade2020learning}. 
In addition, some pre-trained models also take the structural information of the code fragment into account in the initial training phase~\cite{niu2022spt,lin2021traceability}.
For example,  Guo et al.'s GraphCodeBERT ~\cite{hin2022linevd}  to infer different in the data flow of code fragments with the help of graph structures.
DOBF ~\cite{lachaux2021dobf} introduces a novel pre-training objective predicated, explore whether such pre-training can enhance the model's ability to learn the syntactic and structural complexity of source code.
The objective is specifically tailored to address the structural dimension of programming languages.
Concurrently, to enhance the graph-based representation, GraphCodeBERT ~\cite{guo2020graphcodebert} proposed a pre-trained schema to seamlessly insert  graph structure into Transformer-based architectures.
This is achieved through the innovative use of graph-guided masked attention mechanisms, designed to mitigate noise in the data.
The comparative evaluation positions CodeBERT as a baseline standard, which is a very classic pre-trained code model in a series of code-related tasks, including code clone detection and code translation. 
At the same time, UniXcoder as the latest pre-trained code model will also be done as a baseline method.
UniXcoder, a unified cross-modal pre-trained programming language model, is trained on a large amount of code data as well as natural language data~\cite{guo2022unixcoder}. 

Since these pre-trained models have shown superior performance in various code-related tasks, some studies have attempted to use these models for vulnerability detection ~\cite{fu2022linevul,thapa2022transformer,hin2022linevd}. 
However, if these models are directly used for vulnerability detection after fine-tuning with code data, they face the challenge of capturing vulnerability features from long code and complex structure~\cite{zhang2023vulnerability}.
Moreover, due to the nature of code data, the lack of vulnerability semantics information prevents these multi-modal models from taking full advantage of them.
Therefore, we try to supplement the vulnerability semantics in the existing code data to reduce the cost of modeling and searching vulnerability features in complex code data.

\begin{figure*}
\centering
\includegraphics[width=0.98\textwidth]{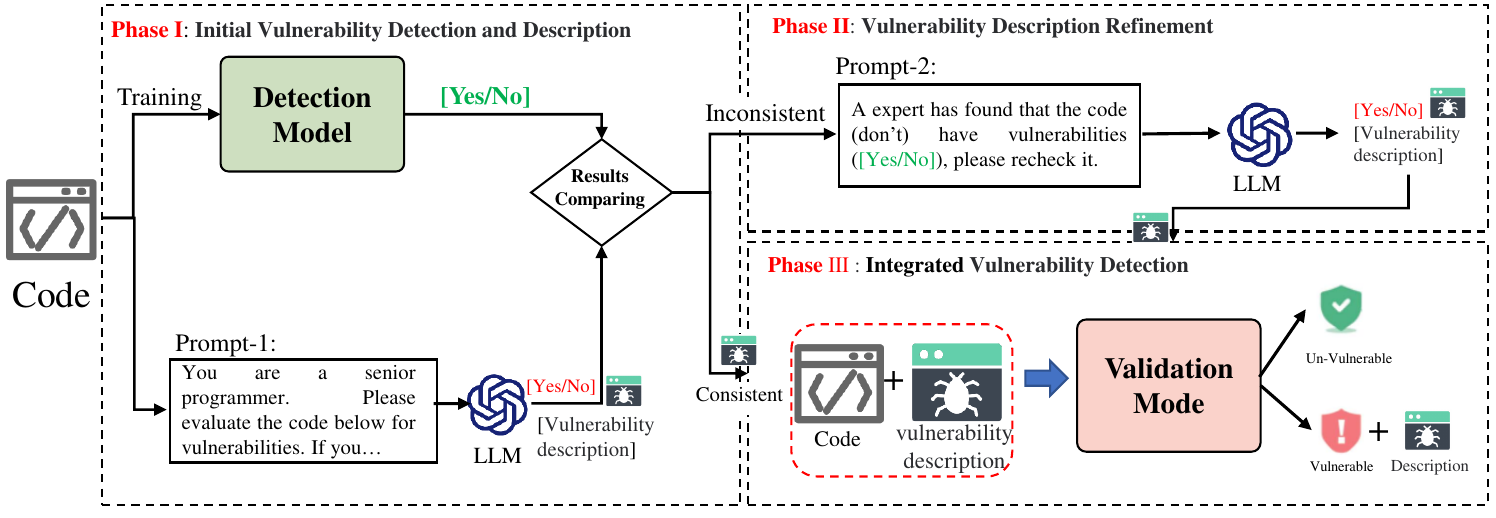}
% \captionsetup{justification=centering}
\caption{The framework of M2CVD, which mainly contains three phase: 1. Initial vulnerability detection; 2. vulnerability description refinement and 3. Integrated vulnerability detection. The detection model uses historical vulnerability data to fine-tune, and then fine-tunes the validation model after the historical vulnerability data is supplemented by vulnerability semantics in phase 1 and phase 2.
% \yxcomment{The inverse of vulnerable is invulnerable but not un-vulnerable.} 
}
\label{fig:main}
\end{figure*}

\section{Approach}
% In this section, we present the architecture of the Multi-model Collaborative Vulnerability Detector (M2CVD).
%
In general, M2CVD requires three models to work together, including code model $f_d$,$f_v$ and LLM $f_c$, and relies on the collaborative interaction of $f_d$ and $f_c$ to assist the enhancement of $f_v$ model in the vulnerability detection task.
The overall framework of M2CVD as shown in Figure \ref{fig:main}, consisting of three phases:

1) Phase I generates preliminary judgments and vulnerability descriptions with the help of $f_c$ and $f_d$.

% %
2) In phase II, the judgments that are inconsistent with $f_c$ and $f_d$ in Phase I  will be judged and described by $f_c$ for the second time.

% %
3) The last phase uses the vulnerability text to enhance the vulnerability detection ability of $f_v$.

In the default configuration of this paper, the code model used by M2CVD is UnixCoder and LLM is ChatGPT 3.5.
% %
% For the convenience of description, in the default case, the pre-trained code model used by M2CVD is UniXcoder
% \footnote{https://github.com/microsoft/CodeBERT/tree/master/UniXcoder}, and the default LLM is chosen to be ChatGPT.
% UniXcoder is an excellent open source code model that is well documented and easily accessible. In RQ4, we will discuss the results of different LLMs and code models choices in detail.
% In general, M2CVD requires three models to work together, including code model $f_d$,$f_v$ and LLM model $f_c$, and relies on the collaborative interaction of $f_d$ and $f_c$ to assist the enhancement of $f_v$ model in the vulnerability detection task.
\subsection{Initial Vulnerability Detection}

% The aim of this phase is to obtain preliminary vulnerability assessments and descriptions from the detection model and LLMs model for code snippets.
%
% In order to execute this phase systematically,we have decomposed it into two steps. 
%

 We use $L = (P, Y) $ to represent the historical vulnerability dataset, where $p_i$ represents a  code snippet in a programming language, and $y_i$ represents vulnerability labels, $0 < i \leq M$.
 The $M$ is the number of code snippet.
 The values of $y_i$ are $0$ or $1$. When $y_i = 0$, it indicates that the code is free of vulnerabilities; conversely, it indicates a vulnerability in the code.

First, we split the vulnerability dataset $L$ into a training set $p_t$ and a validation set $p_v$. In the inference phase, the code to be detected is denoted as $p_e$.
Then, we fine-tune the detection model using $p_t$.
% In this step, the detection models adopted in this paper is UniXcoder~\cite{nguyen2022regvd}. 
According to the methods provided by the existing literature~\cite{steenhoek2023empirical}, the detection model $f_d$ is obtained by fine-tuning on the historical vulnerability data.
% This fine-tuning occurs in a supervised manner, where the model is learned to labeled examples from the dataset $L$, allowing it to adjust its weights to minimize the prediction error for the detection task. 
%  During fine-tuning, the UniXcoder model takes the sequence of code embedding $X = {x_1, x_2, ..., x_n}$ and passes it through additional Transformer layers.
% During training, AdamW is used as an optimizer to select the best model based on the optimal accuracy on the validation set $p_v$.
%

After obtaining the detection model, the vulnerability assessment of the detection model and the vulnerability assessment and description of the LLM are obtained through the following two steps:

 1) Generation of the assessment with detection model. 

We need to use the detection model to complete the preliminary vulnerability assessment for $L$. The specific steps for this are:
\begin{equation}
z_{i}=f_{d}(p_i), 0 < i \leq M
\end{equation}
where $f_{d}$ represents the prediction step of the detection model, and $z_i$ denotes the detection model's assessment of code snippet $p_i$.
If the model determines that $p_i$ contains a vulnerability, then $z_i=0$; otherwise, $z_i=1$. 
%
% It is worth noting that prior to this step, we have already divided the dataset into training, validation, and test sets according to the existing literature~\cite{zhou2019devign}. 
%
% This phase only uses the training set to fine-tune the UniXcoder model.
%
% Subsequently, we select the model that shows the best performance on the validation set to assess vulnerabilities across the entire dataset. 
%
% Finally, the entire dataset including training, validation, and testing is required to perform vulnerability  prediction through the fine-tuned code model.

 2) Generation of the assessment and description with LLMs. 
In this step, we conduct an initial vulnerability assessment and description of $p_t,p_v,p_e$ through an interactive approach using ChatGPT. 
%
% Based on API access restriction rules of OpenAI, the version of ChatGPT employed in this study is ChatGPT 3.5-Turbo-16K~\cite{ChatGPT}.
%
We use the following prompt to obtain LLM's assessment and description:

\renewcommand{\arraystretch}{1.5}
\setlength{\tabcolsep}{10pt}
\begin{table}[H]
% \vspace{-1em}
\small
\centering
% \caption{a }
\begin{tabular}{|p{12.5cm}|}
\hline
\rowcolor{gray!20} % 这一行有20%的灰色阴影'
User:You are a senior programmer. Please evaluate the code below for vulnerabilities. If you believe there are vulnerabilities, reply starting with 'Yes' and briefly explain the issue; otherwise, begin with 'No'. 

Code:  int ff\_get\_wav\_header(AVFormatContext *s, AVIOContext *pb ......
  \\
\hline
 LLM: [Yes][Vulnerability description] or [No] \\

\hline
\end{tabular}
% \vspace{-1em}
\end{table}
Following the above mentioned steps, M2CVD obtains the vulnerability assessment and description form LLMs. The formalized procedure is detailed as follows:
\begin{equation}
c_i,n_i=f_{c}(p_i), 0 < i \leq M
\end{equation}
By analyzing the responses from the LLM, M2CVD obtains a vulnerability assessment $c$.
If the LLM determines the code $p_i$ to be vulnerable and replies with "Yes", then $c_i=0$.
Concurrently, M2CVD records the vulnerability description $n_i$ provided by the LLM. 
If the LLM determines that the code is not vulnerable, 
$c_i=1$ and $n_i$ is set to null.

\subsection{Vulnerability Description Refinement}
Through the process described above, we obtained two vulnerability assessments from the detection models and the LLM, as well as a vulnerability description from the LLM.
In Phase II, we need to further refine the vulnerability assessments and descriptions from the LLM.

In this section involves a comparative analysis of the vulnerability assessments. 
In the case of an inconsistency between assessments, LLMs can be informed of the vulnerability assessment derived from the detection model.
The second interaction enables the LLM to obtain the assessment result of the detection model that fine-tuned based on the historical data, which may enable the LLMs to regenerate its vulnerability assessment and description.
The prompt for Phase II is as follows:

\renewcommand{\arraystretch}{1.5}
\setlength{\tabcolsep}{10pt}
\begin{table}[H]
% \vspace{-1em}
\small
\centering
% \caption{a }
\begin{tabular}{|p{12.5cm}|}
\hline
\rowcolor{gray!20} % 这一行有20%的灰色阴影'
User: You are a senior programmer ...  ...

Code:  int ff\_get\_wav\_header(AVFormatContext *s, AVIOCon ......
  \\
\hline
 LLM: [Yes][Vulnerability description] or [No] \\
 \hline
 \rowcolor{gray!20}
 User: Another expert has found that the code [does not] have vulnerabilities, please recheck it, and If you believe there are vulnerabilities, reply starting with 'Yes' and briefly explain the issue; otherwise, begin with 'No'."
  \\
\hline
 LLM: [Yes][Vulnerability description] or [No] \\

\hline
\end{tabular}
% \vspace{-1em}
\end{table}

Limiting Phase II to code fragments with divergent prediction outcomes can significantly reduce LLM inference time. M2CVD then proceeds to refine the vulnerability descriptions based on this streamlined approach:
\begin{equation}
c_i,n_i=\begin{cases}  c_i,n_i& \text{ if } c_i==z_i \\  f_c(p_i,z_i)& \text{ else } \end{cases}
\end{equation}
When the assessment results from both models are consistent, the interaction with ChatGPT is terminated. This also means that the refinement process does not trigger.
When the models yield inconsistent assessments, i.e., $c_i \ne  z_i$, a second round of interaction is initiated using the aforementioned prompt.
The values of $c_i$ and $n_i$ are updated accordingly.

\subsection{Integrated Vulnerability Detection}
In Phase III, M2CVD leverages the vulnerability assessments and vulnerability descriptions from the LLM to supplement the vulnerability semantics of the vulnerability code.
Initially, we restructure the input dataset $L$.
such that $L_c = (P,C,N,Y) $. The $C$ represents LLM's assessment, while $N$ denotes the LLM's description of the vulnerability in the code segment.
After integrating the vulnerability semantics into the dataset $L_c = (P,C,N,Y) $, the \textbf{validation model $f_v$} was obtained by fine-tuning. 
% \vspace{-1em}
\begin{algorithm}
\caption{M2CVD}
\begin{algorithmic}[1]
\REQUIRE $L = (P,Y)$, code model $f_{d}$,$f_{v}$,LLM:$f_c$ 
% \ENSURE $\hat{y}$
% \\\#Training the detection model
\STATE Split the dataset $L$ into $p_t, p_v, p_e$
\STATE Fine-tune the detection model $f_{d}$ through $p_t$
% \\\#Training the validation model
\STATE The assessment results of detection models on the dataset were calculated: $z_i = f_{d}(p_i)$
\FOR{$i = 1$ to $\text{len}(p_{i})$}
\STATE  $c_i, n_i = f_c(p_{i})$
\IF{$z_i \neq c_i$}
\STATE  $c_i, n_i = f_c(p_{i},z_i)$
\ENDIF
\ENDFOR
\STATE The new data: $p^{'}_{i} = p_{i} + n_i$
\STATE Fine-tune the validation model $f_{v}$ through $p^{'}_{t},p^{'}_{v}$
\\\#Inference phase
\STATE The assessment results of validation model were calculated: $\hat{y} = f_{v}(p^{'}_e)$
\RETURN $\hat{y}$
\end{algorithmic}
\end{algorithm}
% \vspace{-1em}

\textbf{Data Identically Distributed Guarantee.} 
For the fine-tuning process of the validation model, it is crucial to ensure that the training set undergoes both Phase I and II for filling in vulnerability semantics. This approach differs significantly from merely using a vulnerability label for semantic completion via the LLM.
We have found that limiting the enhancement of code vulnerability semantics to only those with identified vulnerabilities in the training set can lead to overfitting in the validation model.
Therefore, in order to ensure that the training set and the validation test data remain identically distributed as much as possible, it is necessary to prohibit directly informing LLM data labels during the vulnerability semantic generation process in Phases I and II.
This necessity occasionally leads ChatGPT to erroneously add semantics to codes perceived as vulnerable.
Nevertheless, maintaining the same data source during both the training and inference phases, while integrating a degree of noise, has effectively increased the robustness of the validation model.

\subsection{Inference Phase}

In the inference phase as shown in algorithmic 1, we follow phases 1 and 2 to get the LLM's vulnerability description $n$ and assessment $c$ for the code to be detected $p_e$. Finally, it was combined as the input of the validation model to complete the vulnerability detection task. The code to be detection: $p_{e}^{'} =<p_e,c,n>$
The formal procedure is articulated as follows:
\begin{equation}
\hat{y_i},n_i=f_{v}{(p_{e}^{'})},0< i\le M
\end{equation}
Here, $\hat{y_i}$ signifies the ultimate assessment result corresponding to $p_e$. $n_i$ is the vulnerability description of the code to be tested. After the M2CVD process, $n_i$ can be used to assist programmers to modify the vulnerability.
The inference of the detection model and LLM for the code under test are performed simultaneously, and usually the inference time of LLM is greater than that of the detection model. Therefore, the inference cost is the time to call the LLM API plus the time with the validation model.
\subsection{Loss Function}
The loss function adopted for the code models training is the cross-entropy loss~\cite{zhou2019devign}, commonly used in classification problems for its effectiveness in penalizing the predicted labels and the actual labels:
\begin{equation}
H(y, \hat{y}) = -y \log(\hat{y}) - (1 - y) \log(1 - \hat{y})
\end{equation}
% This canonical form is used to optimize the detection model and validation model in the M2CVD to ensure the prediction accuracy of the model.
\subsection{Implementation Details}

M2CVD has two processes of model fine-tuning, in which the fine-tuning process of the first evaluation model adopts the best performance parameters reported by the existing code model, specifically referring to\footnote{https://github.com/microsoft/CodeXGLUE/tree/main/Code-Code/Defect-detection}. After completing the vulnerability assessment report, M2CVD implements the second model fine-tuning process, in which epoch is 4, the sequence length takes the maximum length of the base code model used, which is 1024 for uniXcoder by default in this paper, the learning rate is 2e-5, and the bachsize is 12.

\section{Experiments}

\subsection{Datesets}
To evaluate the effectiveness of M2CVD, we employ two datasets from real projects :(1) Devign ~\cite{zhou2019devign}, and (2) Reveal~\cite{chakraborty2021deep}.
The Devign dataset, derived from a graph-based code vulnerability detection study ~\cite{zhou2019devign}, stands as a dataset of function-level C/C++ source code from the well-established open-source projects QEMU and FFmpeg. 
Aligning with the methodology articulated by Li et al. ~\cite{zhou2019devign}, the partitioning of the Devign dataset adheres to a conventional 80:10:10 ratio, demarcating the bounds for training, validation, and testing data, respectively. 
The dataset completes the labeling of vulnerable code by a group of security researchers performing a rigorous two-stage review.
In the task of software vulnerability detection, the REVEAL dataset is a representative dataset, as presented in ~\cite{chakraborty2021deep}. It is a further exploration of data redundancy and unrepresentative class distributions in existing datasets.
As a detection code dataset, REVEAL encompasses source code extracted from two open-source forays: the Linux Debian kernel and Chromium. 
Similar to the real-world situation, this dataset has an imbalanced label distribution, with the number of normal code fragments much larger than the number of vulnerable ones (10:1).
Similarly, in the Reveal dataset, a split ratio of 80:10:10 was set~\cite{ding2024traced}.
%
% \begin{table}[ht]
% \centering
% \caption{Datasets (\#Instances)}
% \label{my-label}
% \begin{tabular}{lccccc}
% \hline
% \textbf{Dataset}            & \textbf{train} & \textbf{dev} & \textbf{test} & \textbf{Vul} & \textbf{Non-Vul} \\ \hline
% \textbf{Devign}             & 21,854           & 2,732         & 2,732          & 10,067        & 12,294            \\
% \textbf{Reveal}             & XXXX           & XXXX         & XXXX          & 1,664         & 16,505            \\ \hline
% \end{tabular}
% \end{table}

%
During the experiment, the proportion of positive and negative samples in the training set, validation set and test set is consistent with the original dataset.

% \begin{table}
% \centering
% \caption{Properties of all the designed test cases.}
% \begin{tabularx}{1cm}{lllXX}
% \hline
% \hline
% Dataset&Non-vul&Train:Test&TrainingData&TestingData\\
% \hline
% D1.1&0.05 & 5\%:95\% & 98,721 &1,774,099 \\
% D1.2&0.10 & 10\%:90\% & 197,440 &1,676,166\\

% \hline
% \hline
% \end{tabularx}
% \end{table}

\subsection{Performance Metrics}
In the process of evaluating the performance of the model, the proposed method includes three metrics widely recognized in the field of software testing and analysis\cite{zhou2019devign}:

\textbf{Precision}: Denoted as the quotient of the sum of true positives and false positives and is a measure of the accuracy of instances that are identified as positive.
Formally, it is defined as: 
\begin{equation}
Precision = \frac{TP}{TP + FP}
\end{equation}
where TP and FP represent the number of true positives and false positives, respectively.

\textbf{Recall}:  Recall evaluates the fraction of actual positives that are correctly identified and is calculated as the fraction of true positives over the sum of true positives and false negatives: 
\begin{equation}
Recall = \frac{TP}{TP + FN}
\end{equation}
where FN signifies the number of false negatives.

\textbf{F1 Score}: The F1 score provides an indicator of the accuracy of the test by combining precision and recall into a single metric by taking their harmonic mean: 
\begin{equation}
F1\ Score = 2 \times \frac{Precision \times Recall}{Precision + Recall}
\end{equation}

\textbf{Accuracy}: This metric reflects the proportion of true positives and true negatives amongst all evaluated instances, thus offering an overall measure of the model’s performance: 
\begin{equation}
Accuracy = \frac{TP + TN}{TP + TN + FN + FP}
\end{equation}
where TN representing the number of true negatives.

REVEAL is an imbalanced dataset, so we emphasize the use of F1 as the evaluation metric~\cite{ding2024traced}. The designed dataset is balanced, so we follow the original benchmark to report the classification accuracy~\cite{zhou2019devign}.

Since the model performance can vary with different random seeds \cite{steenhoek2023empirical}, we used the random seed setting commonly used in existing open source methods, seed=42~\cite{feng2020codebert,ding2024traced}.

\subsection{Baseline Methods}
In our evaluation, we compare M2CVD with seven state-of-the-art methods.

(1) ChatGPT~\cite{ChatGPT}: The GPT series models showcases the capabilities of DL in text generation and processing, albeit not specifically tailored for the domain of software vulnerability detection. 
ChatGPT 3.5 provides the ability to abstract code vulnerabilities at a lower cost.

(2) Devign~\cite{zhou2019devign}: Devign is a graph-based model that uses Gated Graph Recurrent network (GGN) to represent the graph combining AST, CFG, DFG and code sequence of the input code fragment for vulnerability detection.

(3) ReGVD~\cite{nguyen2022regvd}: ReGVD treats the problem as text classification by transforming the source code into a graph structure, using token embedding from GraphCodeBERT~\cite{guo2020graphcodebert}, and applying a mixture of graph-level sum and max-pooling techniques.

% (4) SySeVR~\cite{li2021sysevr}: SySeVR uses code statements, program dependencies, and program slicing as capabilities, and utilizes bidirectional recurrent neural networks to detect vulnerable code fragments.

(4) CodeBERT~\cite{feng2020codebert}: CodeBERT use a pre-trained structure that amalgamates natural language and programming language, facilitating a broad spectrum of coding tasks, including but not limited to code understanding and generation.

(5) CodeT5~\cite{wang2021codet5}: CodeT5, a unified pretrained encoder-decoder Transformer model that better leverages code semantics conveyed by identifiers assigned from developers.

(6) UniXcoder-base~\cite{guo2022unixcoder}: This is a unified code representation model that leverages a Transformer-based architecture. UniXcoder extends the capabilities of models like CodeBERT by incorporating a comprehensive understanding of code syntax and semantics, thereby enhancing the model's performance in coding tasks such as code summarization, translation, and completion.

(7) UniXcoder-base-nine: Continue pre-training unixcoder-base on NL-PL pairs of CodeSearchNet dataset and additional 1.5M NL-PL pairs of C, C++ and C\# programming language. The model can support nine languages: java, ruby, python, php, javascript, go, c, c++ and c\#.

(8) TRACED~\cite{ding2024traced}: TRACED employs an execution-aware pre-training strategy to enhance code models' understanding of dynamic code properties, significantly improving their performance in execution path prediction, runtime variable value prediction, clone retrieval, and vulnerability detection.

Experiment Environment: We implemented M2CVD in Python using Tensorflow. The experiments were performed on a machine containing three NVIDIA GeForce GTX A6000 GPU and two Intel Xeon Gold 6226R 2.90 GHz cpus.

\section{Experiments}
In this section, we conduct extensive experiments to demonstrate
our model’s superiority and analyze the reasons for its effectiveness. Specifically, we aim to answer the following research questions:

\textbf{RQ1:} How effective is M2CVD compared with the state-of-the art baselines on vulnerability detection?

In this RQ, the performance of M2CVD is verified with two real-world vulnerability datasets. Given a code fragment to be detected, M2CVD generates a vulnerability description through a large model, and generates code and vulnerability assessment pairs through collaborative process changes. The performance of M2CVD on the test data is evaluated and compared to the SOTA baseline on two datasets.

\textbf{RQ2:} What are the effects of  vulnerability description refinement for M2CVD?

In this RQ, we verified the effectiveness of the components, which included the comparison between the results of fine-tuning after generating the vulnerability description directly through the large model and the results of fine-tuning after refining the vulnerability description in step 2.

\textbf{RQ3:} What are the effects of hints of code models  for LLMs?

In this RQ, we verify whether the first judgment results of the code model have a positive effect on the LLM. We inform the LLM code model or error judgment information respectively to verify the rationality of the M2CVD process.

\textbf{RQ4:} How well does M2CVD generalize across different code models and LLMs?

In this RQ, we validate the performance between different code models and LLMs combinations on the vulnerability detection task, validating the generality of the M2CVD approach.

\begin{table*}[t]
\centering
\caption{Comparison results for different models on Devign and Reveal datasets. The best result for each metric is highlighted in bold.}
\label{my-label}
\small
\begin{tabularx}{\textwidth}{lXXXXXXXX}
\toprule
\multicolumn{1}{c}{Dataset} & \multicolumn{4}{c}{\textbf{Devign \cite{zhou2019devign}}} & \multicolumn{4}{c}{\textbf{Reveal \cite{chakraborty2021deep}}} \\ \cline{1-9} 
 Models& \textbf{Acc} & \textbf{Recall} & \textbf{Prec} & \textbf{F1} & \textbf{F1} & \textbf{Recall} & \textbf{Prec} & \textbf{Acc} \\ \midrule
ChatGPT 3.5 COT&	49.83&	32.24&	33.00&	30.61&	27.70&	26.34&	30.54&	63.72\\
ChatGPT 4o COT&	53.73&	7.46&	45.94&	4.06&	12.51&	22.17&	97.33&	20.97\\
Devign & 56.89 & 52.50 & 64.67 & 57.59 & 33.91 & 31.55 & 36.65 & 87.49 \\
ReGVD & 61.89 & 48.20 & 60.74 & 53.75 & 23.65 & 14.47 & 64.70 & 90.63 \\
CodeBERT & 63.59 & 41.99 & 66.37 & 51.43   & 35.11 & 25.87 & 54.62 & 90.41\\
UniXcoder-base & 65.77 & 51.55 & 66.42 & 58.05 & 39.47 & 26.31 & 78.94 & 91.90 \\ 
CodeT5-base & 65.04 & 54.26 & 64.12 & 58.78 & 40.56 & 38.16 & 43.28 & 88.79 \\ 
UniXcoder-nine &66.98 & 56.33 & 66.63 & 61.05  & 42.19 & 33.77 & 56.20 & 90.72\\ 
TRACED & 64.42 & 61.27 & 60.03 & 61.05 & 32.66 & 21.49 & 68.05 &91.11 \\ \midrule
M2CVD & 68.33 &57.76 & 68.39 & 62.63 &48.10 & 39.03 & 62.67 &91.55 \\ \bottomrule
\end{tabularx}
% \begin{tablenotes}
*The Reveal dataset is an imbalanced dataset. F1 and recall are the primary metrics.
% \end{tablenotes}
\end{table*}

\subsection{RQ1. Effectiveness of M2CVD }
To answer the first question, we compare M2CVD with the seven baseline methods on the two datasets as shown in table 1. 
We can draw conclusions about the performance of M2CVD compared to the baselines across the evaluated datasets. 
%
% Here is a comprehensive performance analysis for M2CVD:

Table 1 presents the performance of ChatGPT on the vulnerability detection task. It is evident that large-scale language models employ aggressive detection logic. Specifically, in ChatGPT 4o, nearly all code snippets were identified as vulnerable, leading to considerably low F1 scores across both datasets.

M2CVD demonstrates a marked superiority in terms of Accuracy on both datasets. 
In the Devign dataset, M2CVD attains the highest Accuracy of 68.33\%, the highest F1 score of 62.63\% and the highest Precision outperforming all other models. 
This indicates that M2CVD has the most balanced performance in correctly identifying vulnerabilities without being skewed towards over-predicting (which would increase recall but decrease precision) or under-predicting (which would do the opposite).

On the Reveal dataset, since the proportion of negative samples in this dataset is 90, the model with strong fitting performance generally exceeds 90\% on ACC.
In this dataset, people are generally interested in the ability of the model to find positive samples (vulnerability).
For M2CVD, Both Recall and F1 metrics maintain the level of optimal level.
These figures not only show that M2CVD maintains its high performance in different testing conditions but also that it consistently understands and predicts code vulnerabilities with high precision and recall. 
The performance improvement of M2CVD on the reveal dataset is much less than that of Devign, which we believe is caused by  the imbalance of the Reveal dataset, and the vulnerability data is far less than the normal data. This allows ChatGPT to add far less vulnerability semantics to this dataset than to the Devign dataset.

\begin{figure*}[t]
    \centering
    \begin{subfigure}[b]{0.45\textwidth}
        \centering
        \includegraphics[width=\textwidth]{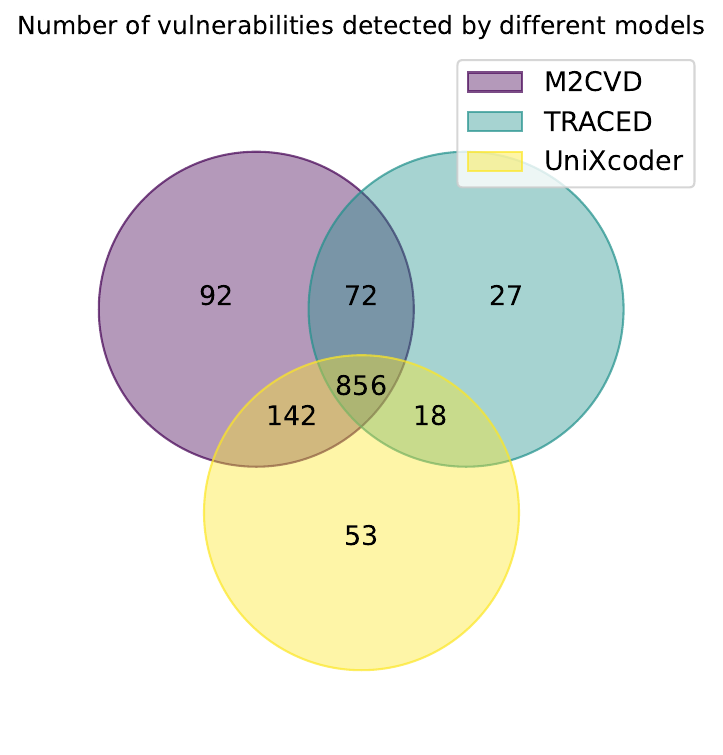}
        \caption{Correct Predictions (Target = 0)}
        \label{fig:correct_predictions_venn}
    \end{subfigure}
    \hfill
    \begin{subfigure}[b]{0.45\textwidth}
        \centering
        \includegraphics[width=\textwidth]{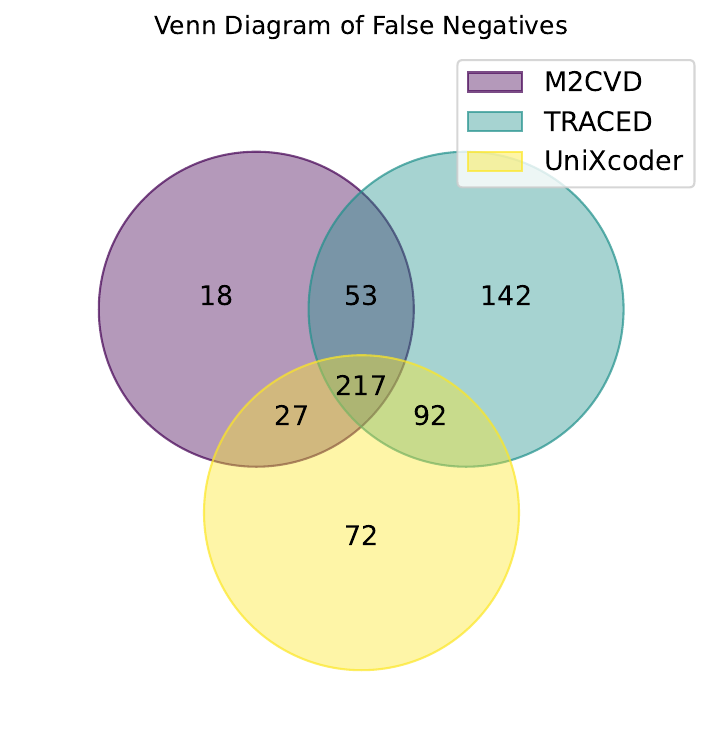}
        \caption{ False Negatives (Predicted = 0, True = 1)}
        \label{fig:false_negatives_venn}
    \end{subfigure}
    \caption{Comparison of the performance of different models in dataset Devign. (a) shows the number of vulnerabilities detected by the M2CVD, TRACED, and UniXcoder models. (b) shows the number of missed vulnerabilities by each model.}
    \label{fig:venn}
\end{figure*}

Figure \ref{fig:venn} presents a comparison of three models (M2CVD, TRACED, and UniXcoder) through Venn diagrams, highlighting their performance in detecting vulnerabilities and false negatives. In Figure \ref{fig:venn}(a), the Venn diagram illustrates the overlap in vulnerabilities correctly detected by the three models: M2CVD independently detects 92 vulnerabilities, outperforming TRACED, which detects 27, and UniXcoder, which detects 53. The three models collectively identify 856 vulnerabilities, indicating a significant overlap and suggesting that they are effective in detecting similar vulnerabilities. M2CVD and UniXcoder jointly detect 142 vulnerabilities, while M2CVD and TRACED jointly detect 72. Figure \ref{fig:venn}(b) depicts the distribution of false negatives among the three models: M2CVD independently produces 18 false negatives, which is significantly lower than the 142 produced by TRACED and 72 by UniXcoder. The three models collectively produce 217 false negatives, indicating some common challenges in detecting certain vulnerabilities. 
In conclusion, the analysis of Figure \ref{fig:venn} demonstrates that the M2CVD model not only excels in detecting a higher number of actual vulnerabilities but also has a lower false negative rate compared to TRACED and UniXcoder. This indicates that the M2CVD model offers superior overall performance in vulnerability detection tasks, making it a valuable tool for applications requiring high accuracy and low false negative rates. Figure \ref{fig:vnn-rv} shows the situation in the Reveal dataset. As shown in Figure Figure \ref{fig:vnn-rv}(a), M2CVD finds more vulnerabilities. However, M2CVD also misses 6 vulnerabilities than UniXcoder due to the complexity of imbalanced data.

In order to better compare the performance of M2CVD, we adopt the UniXcoder-base as the base model, which has the same number of parameters as baseline models such as TRACED and CodeBERT. After the multi-model collaboration process, M2CVD demonstrates superior accuracy and precision over UniXcoder-base, with marked improvements seen in both the Devign and Reveal datasets. 
% While both models show high recall rates, M2CVD slightly surpasses UniXcoder in the Reveal dataset, indicating a better overall detection of relevant instances.
Overall, the performance of M2CVD shows that M2CVD has a more balanced and higher performance in overall performance compared to UniXcoder.

\textbf{Answer to RQ1:}  The performance of M2CVD shows that compared with a single model, M2CVD effectively achieves higher performance in code defect detection tasks under different experimental conditions through a collaborative mechanism.

\begin{table}
\centering
\caption{Model Accuracy Comparison with  different M2CVD configuration }
% \vspace{-1em}
\begin{tabular}{lc}
\toprule
\textbf{Model} & \textbf{Accuracy} \\
\midrule
CodeBERT &63.59\% \\
M2CVD(GPT 3.5+CodeBERT) w/o PII & 65.50\%  \\
M2CVD(GPT 3.5+CodeBERT) &66.10\% \\ 
% \quad \quad \quad \quad \quad \quad \quad \quad \quad -\\
UniXcoder-base & 64.82\%\\
M2CVD(GPT 3.5+UniXcoder-base) w/o PII &67.05\% \\
M2CVD(GPT 3.5+UniXcoder-base) &68.33\% \\
M2CVD(GPT 4o+UniXcoder-nine) w/o PII &64.45\% \\
M2CVD(GPT 4o+UniXcoder-nine) &69.11\% \\

\bottomrule
\end{tabular}
\end{table}

\begin{figure*}[t]
    \centering
    \begin{minipage}[b]{0.45\textwidth}
        \centering
        % \vspace{0.5cm} % Adjust the value to move only the (a) subfigure down
        \includegraphics[width=\textwidth]{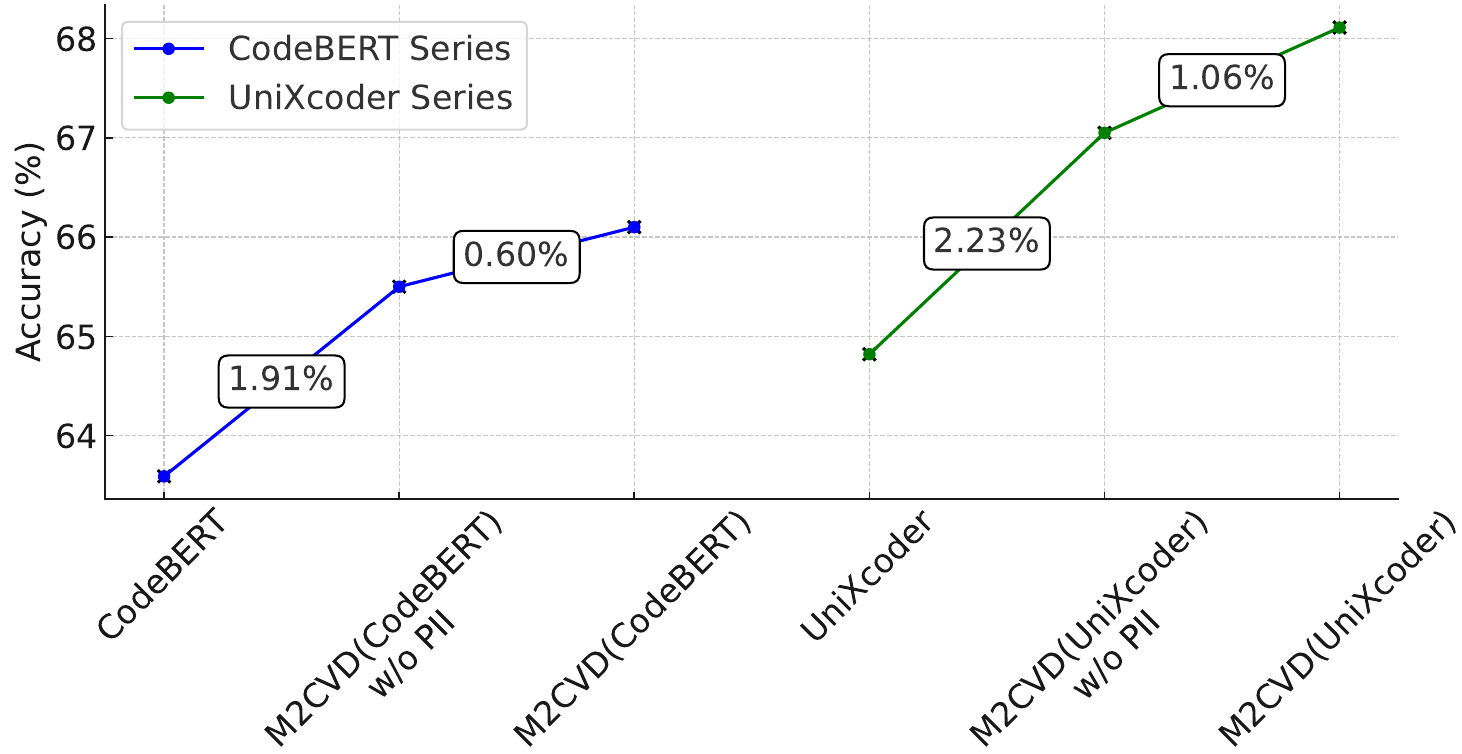}
        \vspace{-0.5cm}
        \caption{Effects of vulnerability description refinement for M2CVD. The figure shows the improvement of the detection effect after CodeBert and UnixCoder pass the two stages of M2CVD, respectively.}
        \label{fig:correct_predictions_venn}
    \end{minipage}
    \hfill
    % \vspace{-1cm}
    \begin{minipage}[b]{0.48\textwidth}
        \centering
        % \vspace{-1cm}
        \includegraphics[width=\textwidth]{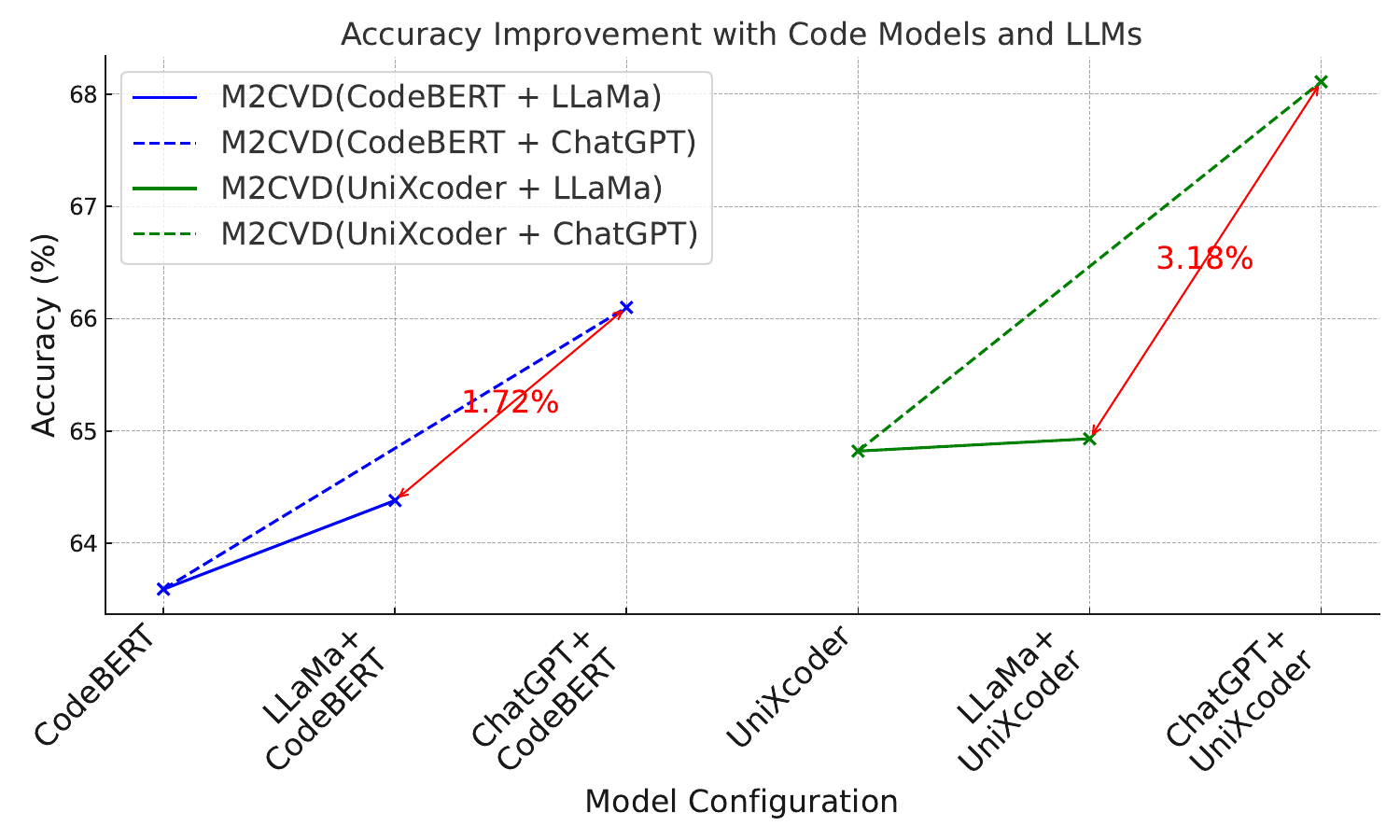}
        \caption{Effects of different code models and LLMs. The figure shows the detection performance improvement of CodeBert and UnixCoder by the vulnerability description models provided by different LLMS, respectively.}
        \label{fig:false_negatives_venn}
    \end{minipage}
    % \caption{Comparison of Model Predictions in dataset Devign}
    \label{fig:model_predictions}
\end{figure*}

\begin{figure*}[t]
    \centering
    \begin{subfigure}[b]{0.45\textwidth}
        \centering
        \includegraphics[width=\textwidth]{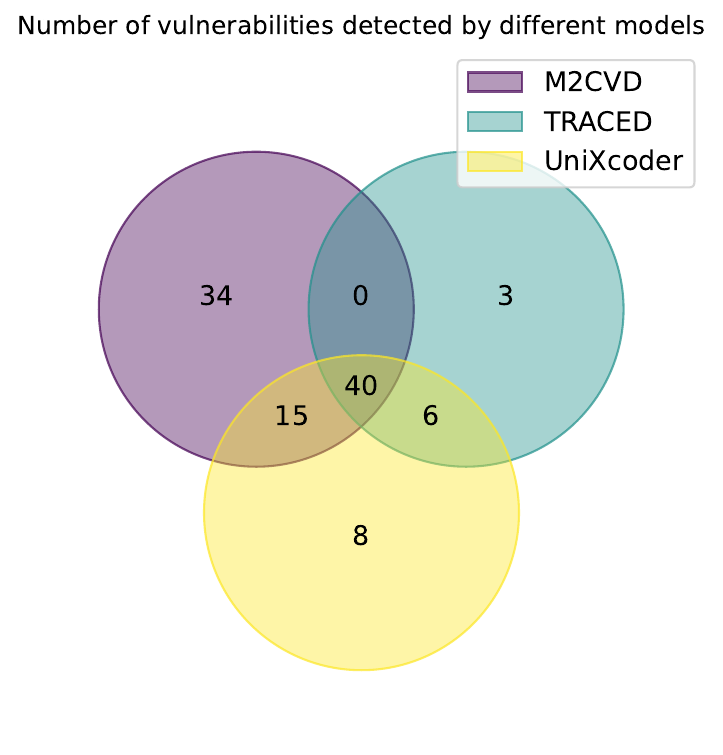}
        \caption{Correct Predictions (Target = 0)}
        \label{fig:correct_predictions_venn}
    \end{subfigure}
    \hfill
    \begin{subfigure}[b]{0.45\textwidth}
        \centering
        \includegraphics[width=\textwidth]{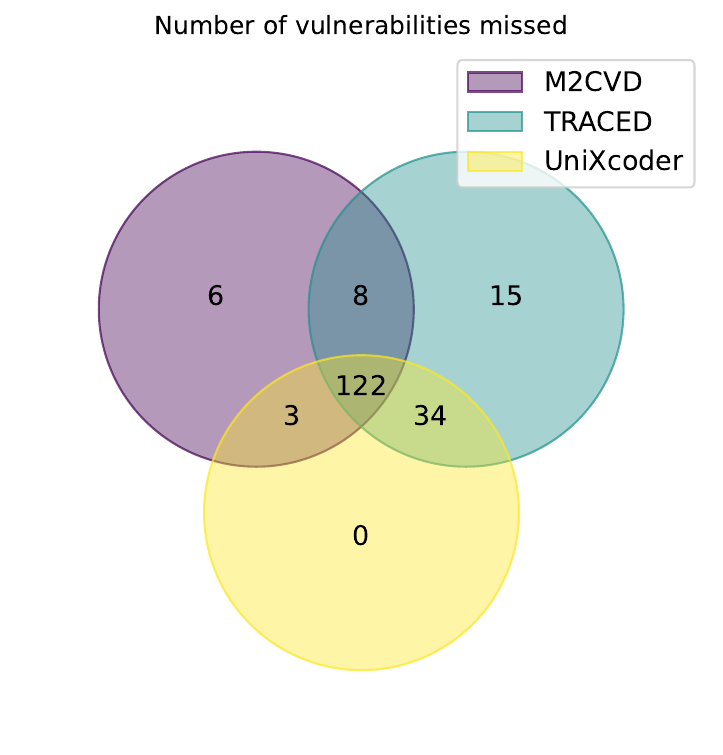}
        \caption{False Negatives (Predicted = 0, True = 1)}
        \label{fig:false_negatives_venn}
    \end{subfigure}
    \caption{Comparison of Model Predictions Result in Reveal dataset}
    \label{fig:vnn-rv}
\end{figure*}

\subsection{RQ2: Effects of vulnerability description refinement for detection performance}
In this section, we elaborate on the implications of Phase II feedback within the M2CVD framework on the performance of code vulnerability detection. 
%
% The aim was to determine whether the process of informing ChatGPT of the code model's judgments after an initial evaluation in M2CVD, and subsequently prompting ChatGPT to re-evaluate the code, was beneficial to model performance.
%

For this purpose, a comparative experiment was established using the Devign dataset with its default partitioning.
The configurations employed in the experiment are as follows:

a) \textbf{CodeBERT}: Utilizes the optimal configuration as reported in existing literature.

b) \textbf{UniXcoder-base}: Also adopts the optimal configuration as documented in existing literature.

c) \textbf{M2CVD(CodeBERT) w/o PII}: This configuration bypasses the M2CVD comparison process, meaning that the LLMs conduct a once assessment without incorporating feedback from the code model's assessments. The description rendered by LLMs is amalgamated with the code, and predictions are made using the CodeBERT model.

d) \textbf{M2CVD(UniXcoder-base) w/o PII}: It omits the M2CVD comparison process. The LLM' first vulnerability detection, when combined with the code, employs the UniXcoder-base model for prediction.

e) \textbf{M2CVD(CodeBERT) }: Use the standard M2CVD process where ChatGPT as the LLMs and CodeBERT as the code model.

f) \textbf{M2CVD(UniXcoder-base)}: Also use the standard M2CVD process, with ChatGPT as the LLMs and UniXcoder-base as the code model.

g) \textbf{M2CVD(GPT 4o + UniXcoder-nine)}: M2CVD experiments were performed on the latest ChatGPT 4o and code model UniXcoder-nine.

The results from the experiment as shown in the table 2, which presents a comparison results of model accuracy under different M2CVD configurations.
The experimental result evidence suggests a improvement in model accuracy when integrating Phase II feedback into the M2CVD framework.
Notably, the M2CVD (CodeBERT) w/o PII outperforms the CodeBERT model by a margin of 1.91\%. 
Similarly, the M2CVD (UniXcoder-base) w/o PII configuration outperforms the  UniXcoder by a margin of 2.23\%.
The enhancements are more pronounced when the comparison includes Phase II, as observed with the M2CVD (CodeBERT) configuration, which incorporates Phase II feedback, outperforms the CodeBERT model by a margin of 2.51\%.  And the M2CVD (UniXcoder-base) configuration with Phase II integration surpasses its UniXcoder-base counterpart by 3.29\%. 

The UniXcoder-nine model is obtained by continuing training on code-natural language pairs based on UniXcoder-base. GPT 4o provides an overly aggressive description of vulnerabilities without the refinement process, and considers almost all of the code to be vulnerable. This leads to the phenomenon that the model overfits on the training set and the vulnerability detection performance decreases.

The results confirm the proposed concept. By integrating vulnerability semantics into code data, we enhance the prediction accuracy of the pre-trained code model in vulnerability detection tasks. Concurrently, these results highlight the efficacy of the vulnerability semantic refinement process within the M2CVD framework. This process significantly boosts the code model's predictive capabilities during the final execution of the code judgment task.

\textbf{Answer to RQ2:} Experiments show that the vulnerability description refinement process of M2CVD can effectively improve the performance of code vulnerability detection.

\subsection{RQ3. Effects of hints of code models for LLMs }
In this section, we elaborate on the implications of Phase II feedback within the M2CVD framework on the performance of the ChatGPT model. 

% Different from RQ2, this section discusses the impact of vulnerability description refinement stage on LLMs without being concerned about the impact on the final result.

The aim is to determine whether informing ChatGPT of the results of the code model, after the initial assessment in M2CVD, benefits ChatGPT's performance for code vulnerability detection. 
For this purpose, we set up a comparative experiment based on the Devign dataset.

The configurations employed in the experiment are as follows:

a) \textbf{ChatGPT}: Get ChatGPT's vulnerability assessment interactively.

b) \textbf{ChatGPT-fewshow}: Get ChatGPT's vulnerability assessment interactively. Two labeled defective code segments and one labeled non-defective code segment are selected as examples using a random selection method.

c) \textbf{ALL“YES” for ChatGPT}: During refinement, we informed ChatGPT of the code model evaluations, but all of them were "YES." This meant that we were mistakenly telling ChatGPT that every fragment of code was vulnerable.

d) \textbf{ALL "NO" for ChatGPT}: During refinement, we informed ChatGPT of the evaluation of code models, but all code models judged "NO". This means that we are wrongly telling ChatGPT that every fragment of code is free of bugs.

e) \textbf{UniXcoder for CodeLLaMa}: Using the standard M2CVD process, the LLMs was chosen as CodeLLaMa.

f) \textbf{UniXcoder for ChatGPT}: Using the standard M2CVD process, the LLMs  was chosen as ChatGPT.\\
where CodeLLaMa-13B is a large language models based on Llama 2. It provides excellent performance among open-source LLMs with long input contexts~\cite{roziere2023code}.

\begin{table}
\centering
\caption{LLMs Accuracy Comparison with  different configurations in Phase II }
% \vspace{-1em}
\begin{tabular}{lc}
\toprule
\textbf{Model} & \textbf{Accuracy} \\
\midrule
 ChatGPT 3.5 COT &45.29\% \\
  ChatGPT 3.5 3-fewshot &45.30\% \\
    ChatGPT 4o COT &53.73\% \\
  CodeLLaMa-13B &49.57\% \\
  UniXcoder-base for CodeLLaMa-13B &52.59\% \\
ALL“YES” for ChatGPT 3.5 & 50.04\% \\
ALL "NO" for ChatGPT 3.5 & 52.12\%  \\
UniXcoder-base for ChatGPT 4o & 56.73\% \\
UniXcoder-base for ChatGPT 3.5 & 57.61\% \\

\bottomrule
\end{tabular}
\end{table}

Based on the experimental data provided in Table 3, we can conclude that refinement step of the M2CVD framework plays a significant role in enhancing the performance of LLMs in detecting code vulnerabilities. 
The experiments reveal several insights: 
The experimental data in Table 5 underscores the impact of the initial feedback given to ChatGPT during Phase II of the M2CVD process. 
Firstly, ChatGPT-fewshot has almost no improvement over ChatGPT. This is due to the many types and complex forms of code defects, and a small number of instances cannot provide effective reference for large models.
When ChatGPT is informed that an expert has judged a piece of code as vulnerable ("YES"), its accuracy in detecting code vulnerabilities increases from 45.29\% to 50.04\%.
This suggests that ChatGPT benefits from additional context. Even by prompting him with insufficient information to check the data again, the detection accuracy can be improved.
Similarly, if ChatGPT is consistently informed that an expert has judged the code as not vulnerable ("NO"), the model's accuracy further improves to 52.12\%. 
%
% This indicates that ChatGPT is also adept at considering expert feedback to dismiss potential false positives, thereby refining its predictions for a more accurate output. 
% The data in Table 5 reveals the efficacy of integrating feedback from the UniXcoder model into the M2CVD process for improving the performance of LLMs like ChatGPT in detecting code vulnerabilities.
%
When the evaluation results of the UniXcoder model fine-tuned on the dataset are provided for ChatGPT, the accuracy of vulnerability detection rises to 57.61\%.
This suggests that the UniXcoder model encapsulates the dataset's inherent logic effectively and can guide the LLM (ChatGPT) towards more accurate evaluations. On the CodeLLaMa model, we observe the same phenomenon, increasing the accuracy from 49\% to 52\%.

The accuracy of the basic ChatGPT model is 45.29\%, while after the refinement process of M2CVD, ChatGPT shows significant performance improvement.
This suggests that from a specialized code model, which carries insights from its fine-tuning process, is crucial in helping LLMs better understand and evaluate the code in question.

\textbf{Answer to RQ3:} The strategy of informing LLMs with the insights from code models that are attuned to the specific dataset logic not only improves the performance but also highlights the potential of collaborative learning systems in code vulnerability detection.
This process effectively enhances the accuracy of LLMs in the code vulnerability detection task within the framework of M2CVD, and also enhances the accuracy of LLMs in adding vulnerability semantics.
  
% \vspace{-1em}

% \section{Disucssion}
% \begin{table}
% \centering
% \caption{Model Accuracy Comparison with  different code models and LLMs working together }
% \begin{tabular}{lc}
% \toprule
% \textbf{Model} & \textbf{Accuracy} \\
% \midrule
% CodeBERT & 63.59\%  \\
% CodeLLaMa-13B+CodeBERT &64.38\% \\
% ChatGPT 3.5+CodeBERT & 66.10\%  \\
% UniXcoder-base& 64.82\%  \\
% CodeLLaMa-13B+UniXcoder-base& 64.93\% \\
% ChatGPT 3.5+UniXcoder-base & 68.11\% \\
% M2CVD(ChatGPT 4o+UniXcoder-nine) &69.11\% \\
% \bottomrule
% \end{tabular}
% \end{table}

\subsection{RQ4. Effects of different pre-trained code models and LLMs working together}
In this section, we elaborate on the impact of different code models and LLMs on the performance of M2CVD vulnerability prediction.
The aim is to judge the impact of the choice of code model and LLMs during M2CVD on the final performance. To this end, we set up comparative experiments using the Devign dataset and its default partition.

The configurations employed in the experiment are as follows:

a) \textbf{CodeBERT}: Utilizes the optimal configuration as reported in existing literature.

b) \textbf{UniXcoder}: Adopts the optimal configuration as documented in existing literature.

c) \textbf{CodeLLaMa-13B+CodeBERT}: Also use the standard M2CVD process, with CodeLLaMa serving as the LLMs and CodeBERT as the code model.

d) \textbf{CodeLLaMa-13B+UniXcoder}: Also use the standard M2CVD process, with CodeLLaMa serving as the LLMs and UniXcoder as the code model.

e) \textbf{ChatGPT+CodeBERT}: Also use the standard M2CVD process, with ChatGPT serving as the LLMs and CodeBERT as the code model.

f) \textbf{ChatGPT+UniXcoder}: Also use the standard M2CVD process, with ChatGPT serving as the LLMs and UniXcoder as the code model.

The result from the experiment presented in Table 4 provides experimental result about the impact of combining different code models and LLMs within the M2CVD framework for predicting code vulnerabilities. 
The standalone models, CodeBERT and UniXcoder, establish a baseline with accuracy of 63.59\% and 64.82\%, respectively. 
The combination of CodeBERT with CodeLLaMa-13B results in a slight accuracy increase, reaching 64.38\%. When UniXcoder is paired with CodeLLaMa-13B, there is a more noticeable improvement, with the accuracy climbing to 64.93\%.
These figures serve as a benchmark to assess the added value of integrating LLMs with code models.
More substantial gains in accuracy are observed when ChatGPT is introduced to the mix. 
ChatGPT paired with CodeBERT yields an accuracy of 66.10\%, while its combination with UniXcoder tops the table at 68.11\%. 
The result from the experiment presented in fig 4 provides a more intuitive result about the impact of combining different code models and LLMs within the M2CVD framework for predicting code vulnerabilities. 

This indicates that the ChatGPT model significantly enhances the performance of both code models, with the ChatGPT+UniXcoder configuration proving to be the most effective partnership in this experiment.

\begin{table}
\centering
\caption{Model Accuracy Comparison with different code models and LLMs working together}
\begin{tabular}{lcc}
\toprule
\textbf{Model} & \textbf{Parameters} & \textbf{Accuracy} \\
\midrule
CodeBERT & 125M & 63.59\%  \\
CodeLLaMa-13B+CodeBERT & 13B+125M & 64.38\% \\
ChatGPT 3.5+CodeBERT & -+125M & 66.10\%  \\
UniXcoder-base & 223M & 64.82\%  \\
CodeLLaMa-13B+UniXcoder-base & 13B+223M & 64.93\% \\
ChatGPT 3.5+UniXcoder-base & -+223M & 68.33\% \\
ChatGPT 4o+UniXcoder-base & -+223M & 68.88\% \\
ChatGPT 4o+UniXcoder-nine & -+223M & 69.11\% \\
\bottomrule
\end{tabular}
\end{table}

\textbf{Answer to RQ4:}
The synergy between LLMs and code models in the M2CVD framework significantly enhances the precision of detecting code vulnerabilities.
Specifically, the more superior performing LLMs and code models contributes to a more pronounced accuracy improvement in the M2CVD synergy mechanism.

% \vspace{-1em}

\section{Case Study}
\begin{figure}
\centering
\includegraphics[width=0.85\textwidth]{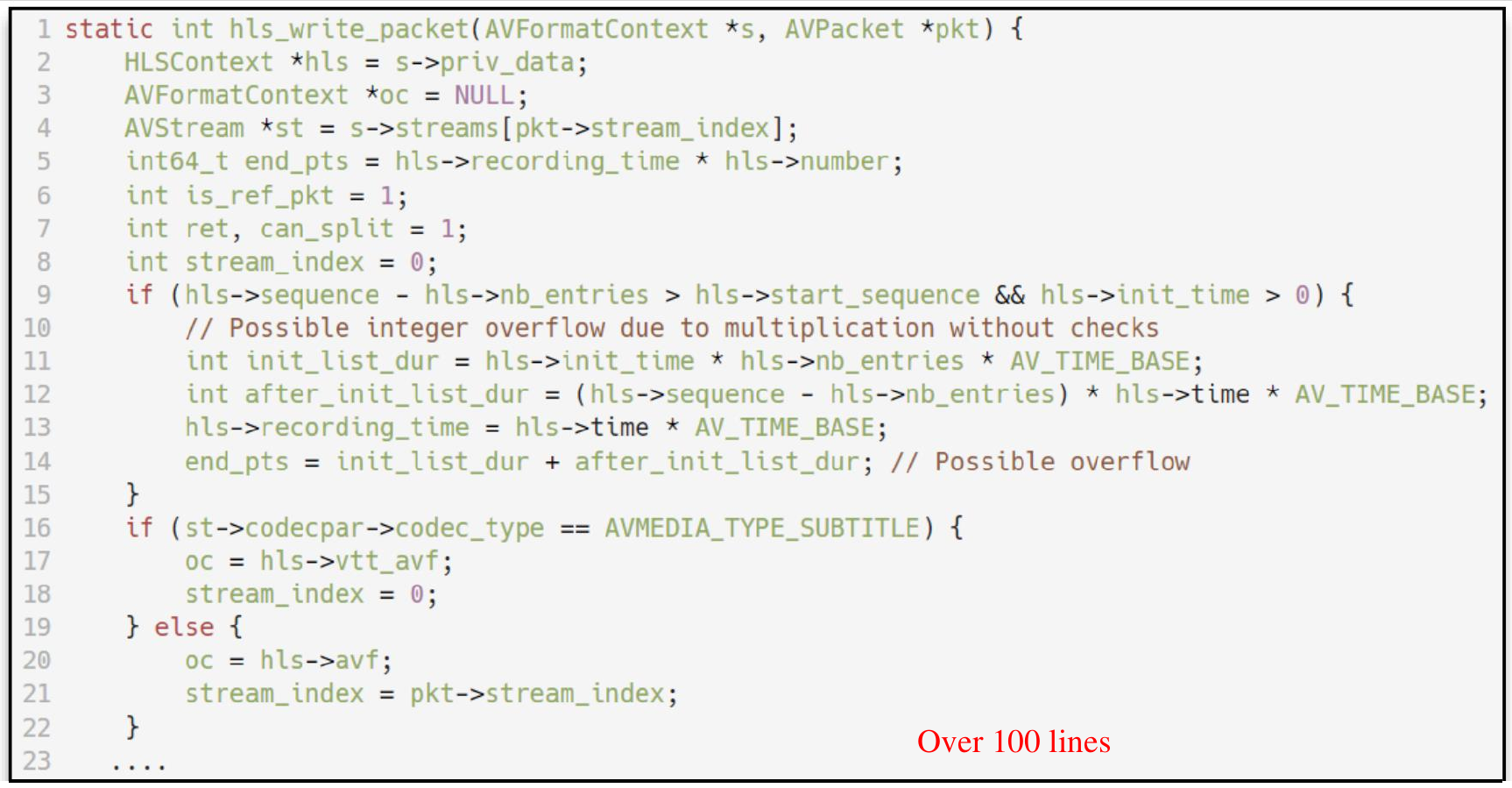}
% \includegraphics[width=\linewidth]{pdf/demo.pdf}
% \vspace{-1em}
% \captionsetup{justification=centering}
\caption{A fragment from the Devign dataset with a code vulnerability. The vulnerability is in a few lines in this very long code.}
\label{fig:code}
% \vspace{-2em}
\end{figure}

In this section, we show instances of LLMs generate code vulnerability semantics. This is the core idea of the M2CVD method.
Figure \ref{fig:code} presents a fragment of vulnerable code from the Devign dataset. This function spans over 100 lines, yet the vulnerability is concealed within just a few of them. 
Traditional vulnerability detection models dissect this code into tokens to learn the vulnerability features.   
However, these features tend to be obfuscated by a large number of no-vulnerability codes, making the learning process challenging.

\begin{figure}
\centering
\includegraphics[width=0.85\textwidth]{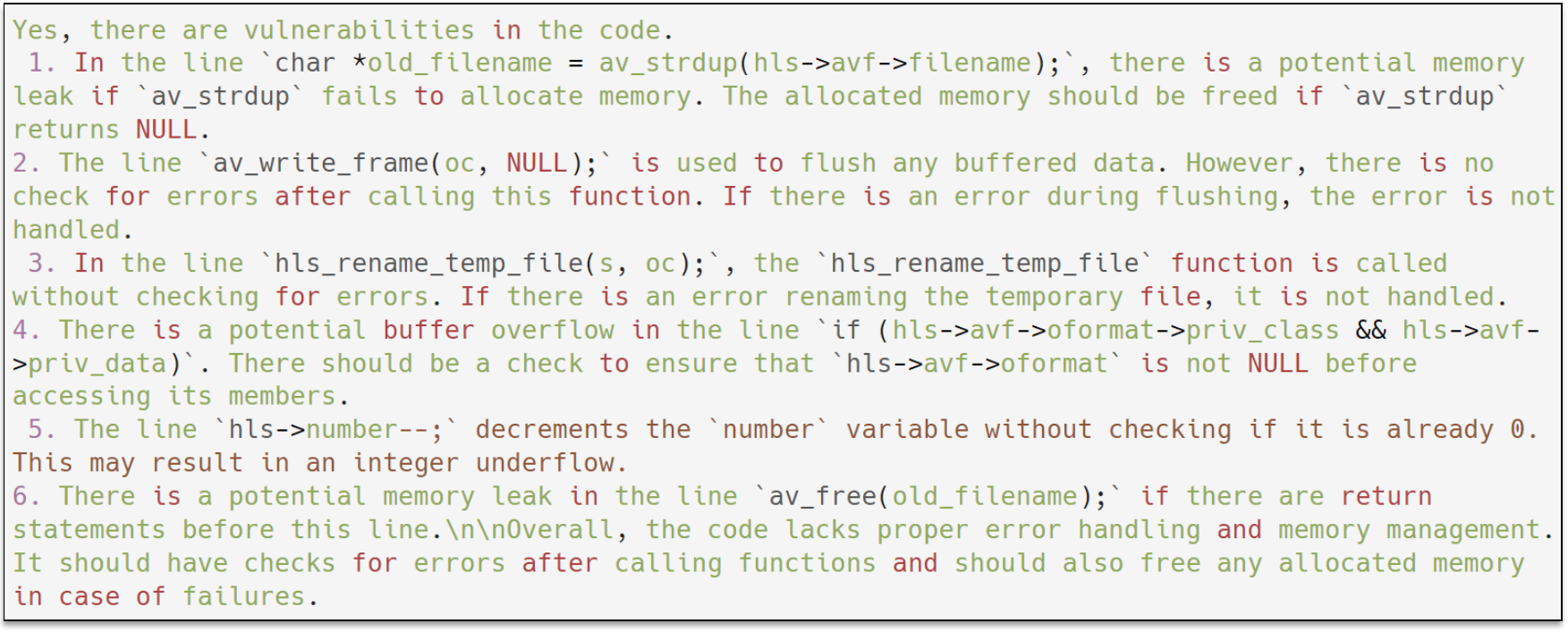}
% \includegraphics[width=\linewidth]{pdf/demo2.pdf}
% \captionsetup{justification=centering}
% \vspace{-1em}
\caption{ChatGPT's abstract representation of the vulnerability in this code fragment(fig.2).}
\label{fig:nl}
% \vspace{-2em}
\end{figure}

Figure \ref{fig:nl} provides a natural language description of the vulnerability present in this code by LLMs.
This model condenses the risky elements of the code into a succinct natural language summary.
This approach offers two significant advantages.
Firstly, the simplification of features substantially eases the code model's learning process, focusing on specific keywords related to the vulnerabilities. 
Secondly, it provides a more consistent representation of vulnerabilities. Similar vulnerabilities are expressed diversely across different code segments, leading to a vast feature space for vulnerability characteristics.
Language models alleviate this issue by standardizing the same vulnerabilities into a unified natural language description, thereby streamlining the learning and identification process.

\section{Discussion}
In this section, we discuss the design of prompt in M2CVD, the reason for version selection of ChatGPT model.

\textbf{Design of prompt.} In the collaborative process of the M2CVD method, two sentences of intuition-based prompt are used to complete the interaction with ChatGPT.
In the prompt used by M2CVD, we followed the experience provided in existing research, setting roles for LLMs and providing task contexts.
Existing literature acknowledges the impact that varying prompts can have on the outcomes yielded by LLMs, with techniques such as Chain-of-Thought~\cite{wei2022chain}. 
However, the focus of this study is mainly to explore the feasibility of multi-model collaboration rather than optimization techniques of prompt, which is a concern more related to the field of prompt technology.
Although the prompts employed within M2CVD may not represent the zenith of optimization, their application has resulted in a significant enhancement of performance in the code vulnerability detection tasks, underscoring the efficacy of the multi-model collaborative approach.

\begin{table}[ht]
\centering
\renewcommand{\arraystretch}{1.3}
\begin{tabular}{|>{\centering\arraybackslash}p{4cm}|>{\centering\arraybackslash}c|>{\centering\arraybackslash}p{4cm}|>{\centering\arraybackslash}c|}
\hline
\textbf{ChatGPT 4o} & \textbf{Count} & \textbf{ChatGPT 3.5-turbo} & \textbf{Count} \\
\hline
Memory Management Issues & 21 & Buffer Overflow & 5 \\
\hline
Input Validation Issues & 18 & Memory Leaks & 5 \\
\hline
Boundary and Overflow Issues & 16 & Improper Error Handling & 5 \\
\hline
Error Handling Issues & 10 & Integer Overflow & 5 \\
\hline
Concurrency and Synchronization Issues & 4 & Null Pointer Dereference & 3 \\
\hline
\end{tabular}
\caption{Top 5 Vulnerability Types from ChatGPT 3.5 and ChatGPT 4o}
\label{table:top_vulnerabilities_comparison}
\end{table}

\textbf{Version of ChatGPT}. 
In the latest release, ChatGPT 4o offers enhanced generation and understanding capabilities compared to ChatGPT 3.5. However, the relatively high usage fees associated with ChatGPT 4o make it impractical for generating vulnerability semantics for tens of thousands of code fragments. On the other hand, ChatGPT 3.5 has more lenient access policies and pricing, making it a more feasible option for large-scale tasks.

We performed M2CVD on the Devign dataset with different ChatGPT and UniXcoder-base. 
We conducted experiments using M2CVD on the Devign dataset, comparing different versions of ChatGPT and UniXcoder-base. The cost of using ChatGPT 4o for defect detection on the full dataset amounted to approximately \$1200. Despite the higher cost, the experimental results indicated that the vulnerability descriptions generated by ChatGPT 4o were not significantly better than those generated by ChatGPT 3.5. Consequently, ChatGPT 3.5 was selected as the default LLM version for M2CVD.

Our sampling analysis on the Devign dataset revealed that ChatGPT 4o reported 98\% of its code as vulnerable due to its overly strict vulnerability definition, resulting in an F1 score of only 8.16\%. After applying the M2CVD process, the F1 score improved to 19.23\%.

Table 5 summarizes the vulnerability reporting by GPT 4o. The high false positive rate of the GPT 4o model can be attributed to several factors. The detector lacks global context information when analyzing code snippets, relying solely on the snippet itself for judgment. This limitation can lead to false positives, such as failing to recognize pre-initialized variables or pre-performed bounds checks in the calling function.
The detector's strict criteria flag even minor potential issues as vulnerabilities, like uninitialized variables or unchecked pointers, even when they are safe within the specific context. The detector is highly sensitive to boundary checking and input validation, resulting in numerous unnecessary warnings even in scenarios where out-of-bounds access is impossible. 
By assuming the worst-case scenario, the detector enhances code robustness in some instances but often imposes excessive error-handling logic that is not required in most real-world applications.

ChatGPT3.5's vulnerability determination logic is not that aggressive. GPT 3.5 gives 70 defect judgments and 30 non-defect judgments for 100 sampled data. The accuracy is 53\% and the F1 is 38.96\%. After the process of M2CVD, ACC is increased to 55\%, and F1 is greatly increased to 50.55\%. GPT 3.5 gives 56 vulnerability judgments, which tend to follow the distribution of this data set. Considering the sampling experiment results, ChatGPT 3.5 is used as the default LLM in this method.

\section{Conclusion}
In this paper, we introduce M2CVD, a novel method designed to address the challenge of software vulnerability detection by harnessing the combined strengths of pre-trained code models and large language models.
The M2CVD integrates the language models such as ChatGPT and code models like UniXcoder, to create a collaboration process capable of detecting vulnerabilities with high accuracy.
Empirical evaluations conducted on the REVEAL and Devign datasets have demonstrated the effectiveness of M2CVD, showcasing its superior performance in detecting code vulnerabilities compared to existing benchmarks.
The results of this research not only confirm the viability of M2CVD as a high-fidelity detection system but also underscore the potential of model synergy in enhancing the capabilities of automated vulnerability detection mechanisms.
In essence, M2CVD demonstrates the potential to exploit the ability of different models to work together, providing a new idea for future research in automated software vulnerability detection and a scalable and effective solution for protecting software systems from changing threats.
% In future work, we expect to further explore how to reduce the impact of complex forms in the code on vulnerability detection, hoping to achieve higher accuracy vulnerability detection by reducing the complexity of the code.

\begin{acks}
To Robert, for the bagels and explaining CMYK and color spaces.
\end{acks}

\bibliographystyle{ACM-Reference-Format}
\bibliography{sample-base}

\end{document}